# Broadband ferromagnetic resonance in Ni-Mn-Ga single crystal

Luděk Kraus[1*], Denys Musiienko[1], Martin Kempa[1], Jaroslav Čapek[1]

[1] Institute of Physics of the Czech Academy of Sciences, Na Slovance 1999/2, CZ-18221, Prague 8, Czech Republic

*corresponding author's email: kraus@fzu.cz

**Abstract**

We present a broadband ferromagnetic resonance study in single crystalline $Ni_{50}Mn_{28.1}Ga_{21.9}$ in the temperature range from room temperature to 120 °C in which the transformation from martensite to austenite phase takes place. Our measurements demonstrate that a large change (an order of magnitude) in the magnetocrystalline anisotropy at the martensitic phase transformation results in a sharp change of the resonance magnetic field. In a single variant martensite phase, the resonance fields satisfy the Kittel's resonance condition for a thin film with the gyromagnetic factor $g = 2.0$. With the magnetic field parallel to the easy c-axis of the single variant martensite, the resonance is observed only for frequencies larger than 22 GHz. For the multivariant martensite case, the magnetic coupling between the twin variants can be taken into account for the satisfactory Kittel's fit. We observe a weak magnetocrystalline anisotropy in the austenite phase, just above the reverse martensite transformation, comparable to the previous reports based on different magnetic measurements.

## 1. Introduction

Ferromagnetic resonance (FMR) is a unique technique for the measurement of magnetic properties of ferromagnetic materials. It is applied to a range of materials from bulk ferromagnetic materials to nano-scale magnetic thin films. The resonance measurements are used to determine fundamental properties of the studied magnetic material, such as magnetocrystalline anisotropy, g-factor, exchange constant, Gilbert damping, etc. in single crystals or magnetic interactions between layers and components in multilayered nanostructured films and foils [1,2]. Ni-Mn-Ga Heusler alloys are a subclass of traditional shape memory materials that exhibit strains and shape change not only in response to the increase of temperature (reverse martensite transformation) but also when exposed to relatively low (< 5 kOe) external magnetic field (twin mobility in martensite phase) [3,4]. These alloys, called magnetic shape memory (MSM) alloys, are mainly known for large reversible magnetic-field-induced strain (MFIS) of up to 12% while staying in their martensitic phases (10M, 14M and NM) [5–7]. FMR has been used to study the angular dependence of the resonance field in MSM alloys providing additional information about their magnetic anisotropy, the main property that defines the magnetic and thermo-magnetic effects in these materials. FMR in epitaxially grown Ni-Mn-Ga thin films, where the twin structures are confined by the rigid substrate, has been thoroughly investigated [8–11]. It was found that the interaction of the twin variants leads to unusual magnetic anisotropy [11]. On the contrary only a few papers can be found on FMR in single crystal magnetic shape memory alloys [12–14]. After the martensitic transformation, MSM single crystals exhibit a random twin structure, accommodating the stresses during the transformation from the austenite. Therefore, the interpretation of FMR experiments for MSM alloys is more complicated in comparison with classic single crystalline magnetic materials. The MSM effect is characterized by a twin boundary motion in an applied external magnetic field, and particular care should be taken when preparing measurements for the unstrained single crystals and the corresponding interpretation of the experimental data.

The temperature and angular dependences of resonance curves were previously measured on bulk Ni-Mn-Ga single crystals at X-band microwave frequency [12,13]. Using the Smit and Beljers resonance condition, from [14], for angular dependence of the resonance field the anisotropy constants $K_1$ and $K_2$ of the martensitic phase were determined. These values substantially differ from the anisotropy constants determined from the magnetization measurements [15,16]. This discrepancy was probably caused by a wrong assumption that the crystallographic structure of the sample had not changed when the field direction varied.

The authors have previously conducted FMR measurements in the temperature range of 20 to 150 °C for a Ni-Mn-Ga single crystalline sample at Q-band frequency for two different orientations of applied field (along (100) and (010) directions of the parental austenite) [17]. Several resonance peaks were observed depending on the actual twin structure: starting from one peak for the single variant up to five peaks for a 3-variant structure. Three of the five peaks were attributed to the usual ferromagnetic (acoustic) resonance modes and the remaining two were explained as the exchange (optical) modes in the two sublattice model for ferrimagnetic materials [18]. Unfortunately, the exchange coupling between the two sublattices obtained from the experiment was about three orders of magnitude smaller than the theoretical value derived from the first-principles calculations. Moreover, the exchange mode was not observed for the single variant case.

Here we present and discuss the broadband ferromagnetic resonance measurements in a Ni-Mn-Ga single crystal in the temperature range from room temperature (RT) to 120 °C, covering both the martensitic to austenitic and the reverse phase transformations of the alloy.

## 2. Experimental

*2.1 Sample preparation and characterization*

The master alloy bar, with a nominal composition $Ni_{50}Mn_{28.5}Ga_{21.5}$, was prepared from pure raw elements by induction melting (Balzers VSG 02) in an $Al_2O_3$ crucible in atmosphere of pure Ar (> 99.9%). After melting of the whole batch, the melt was homogenized for approximately 5 min., subsequently, the furnace was evacuated to remove dissolved gases and prevent pore formation during the solidification. The degassed melt was then cast into a hollow Cu mold with the internal diameter of 11 mm. The master alloy bar was then placed inside a sapphire crucible with an internal diameter of 12 mm for the resolidification in the Optical floating zone system FZ-T-12000-X-VI-VP by Crystal Systems Corporation. The bar inside the sapphire crucible was slowly heated by radiation generated by four Xe lamps. The radiation was focused in a narrow zone, located approximately 10 mm from the bottom of the bar. After the bar was melted within the hot zone, the heating power ramp was stopped, and the bar was translated at a constant speed of 5 mm/h downwards. The process was done in a stationary atmosphere of pure Ar (> 99.95%) and sample rotation of 5 RPM to achieve homogeneous temperature distribution in the hot zone. After the resolidification was completed the power of Xe lamps was slowly decreased to prevent defect formation. A significant portion of the resolidified bar was found to consist of two single crystalline grains. The grains were then oriented by using the PANalytical X'Pert PRO diffractometer with Co anode ($\lambda(K\alpha_1)$ = 0.178901 nm, $\lambda(K\alpha_2)$ = 0.179290 nm) as the X-ray source, and then cut in a Princeton Scientific Corp. WS-25 high-precision wire saw into cuboid samples with their faces parallel to $<100>_A$ planes of the parent austenite phase. X-ray diffraction revealed the monoclinic 10M structure with lattice parameters a = 0.5968 nm, b = 0.5944 nm, c = 0.5593 nm and $\gamma$ = 90.4°. Chemical composition of $Ni_{50}Mn_{28.1}Ga_{21.9}$ of the grown single crystalline samples was measured in an X-ray fluorescence spectrometer, EDAX® AMETEK® ORBIS PC, equipped with a Rh anode tube ($E(K\alpha)$ = 20.216 keV) and Apollo XRF ML-50 EDS detector.

A cuboid single crystalline sample with dimensions 3.5 × 2.5 × 1.1 mm$^3$ was chosen for the FMR measurements presented in this work. For simplicity we chose the coordinate system ($x$, $y$, $z$) with $x$ along the long, $y$ along the shorter and $z$ along the shortest sample axis (see Fig. 1). The sample surfaces were mechanically grinded on an abrasive paper (down to 4000 grit) and then mechanically polished on a cloth surface by consequently using diamond slurries with the grain sizes of 3 μm and 1 μm. The sample was then electro-polished in an electrolyte mixture of one part of 60% $HNO_3$ and three parts of ethanol p.a. at a temperature of −20 °C.

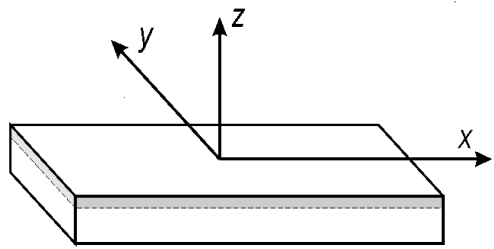

Fig. 1. Schematic view of the sample and the orientation of coordinate axes. The layer on the surface indicates the skin depth, where the resonance takes place.

*2.2 Static magnetic measurements*

Saturation magnetization $M_s$ in the temperature range from RT to 150 °C was determined from the DC-magnetization measurement performed in MicroSense VSM EZ7 system in a constant magnetic field of 12 kOe. To avoid the possible magnetic reorientation and corresponding changes of sample dimensions in the martensitic state, the demagnetizing factors were measured in the austenite phase at 80 °C. The factors $D_x$ = 0.162 and $D_y$ = 0.215 were obtained from the initial susceptibilities measured with magnetic field applied along x and y axis of the sample, respectively. The possible changes of the demagnetizing factors on transition to the martensitic state and magnetic reorientations during measurements were, for simplicity, neglected. The effective anisotropy field of martensite $H_K$ of about 6.95 kOe at 22 °C was obtained from the hysteresis loop measured with the field applied along the hard anisotropy axis of the sample.

*2.3 Broadband ferromagnetic resonance*

Broadband ferromagnetic resonance was done on a home-made equipment schematically shown on Fig. 2. The sample is placed on a U-shaped co-planar waveguide (CPW). It is insulated from the waveguide by a thin Teflon foil. The microwave power, typically 10 dBm, is supplied from one port of the vector network analyzer Keysight P5008A. Magnetic field of Varian 8-inch electromagnet is modulated with the frequency of 25 Hz by a pair of coils fed by an amplified reference signal of Stanford Research SR830 lock-in amplifier. The transmitted signal, detected by the microwave diode Krytar 604B, is amplified by the lock-in amplifier. The temperature of the sample was controlled by a planar heater placed below the co-planar waveguide. A small Pt thermo-resistor in contact with the sample was used to record the temperature of the sample. The whole system is controlled by a LabView program. The resonance curves are measured at different constant microwave frequencies up to 42 GHz. The measured signal, which is proportional to the field derivative $dP/dH$ of microwave power $P$ transmitted through the microwave circuit, is, unfortunately, affected by the spurious signal due to co-planar waveguide and the microwave connectors, which are slightly magnetic. The spurious signal measured from an empty circuit is therefore mathematically subtracted from the measured curves during the data processing.

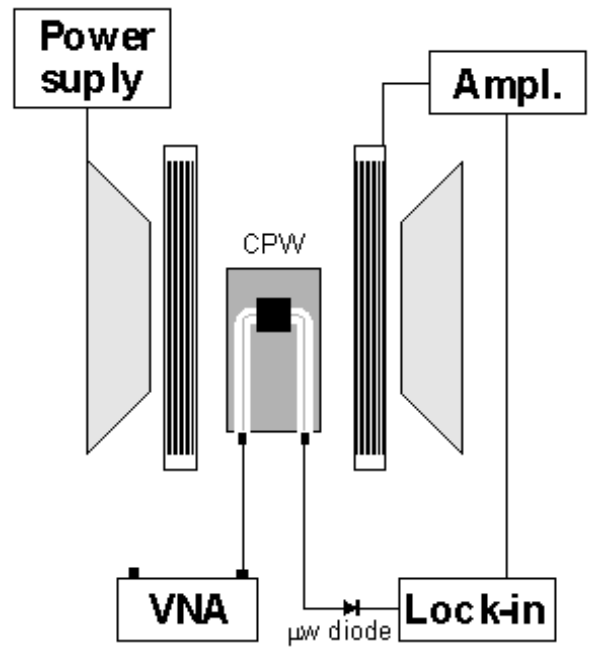

Fig. 2. Principial scheme of BB-FMR apparatus.

3. Results

For the evaluation of resonance data, the modified Kittel's resonance conditions for tangentially magnetized thin films [19] are used. Because of skin effect the ferromagnetic resonance in bulk metallic samples take place in a thin layer at the surface, as schematically shown on Fig. 1. It was shown that in spite of nonuniform microwave field in the skin layer the equations, derived for uniform excitation field, well describe the observed resonance field (see [19] and references cited within).

Four types of resonance measurements were performed. Before each set of measurements, the sample was field-cooled at 15 kOe from 140 °C to room temperature with external magnetic field applied either along the long ($x$) or the short ($y$) axes of the sample. Then the sample was either left in the same orientation for the duration of the measurement or rotated by 90° about the axis perpendicular to the $xy$-plane and during the FMR experiment the external magnetic field was swept along x- or y-axis. Schematic representation of the sample orientation is shown as the inset in Fig. 3a.

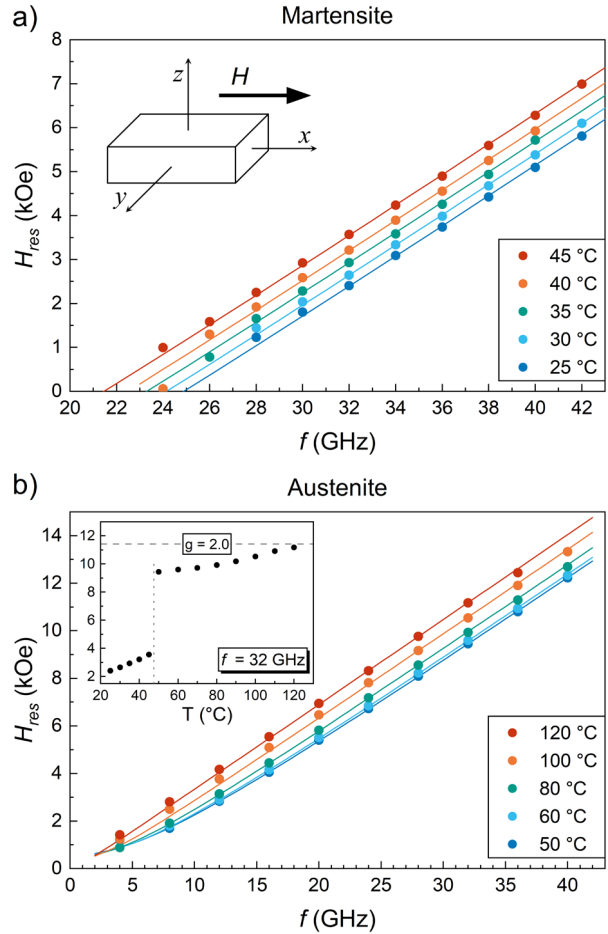

Fig. 3. Resonance fields as function of frequency measured after field cooling along x-axis and measured with field in the same direction, a) martensite, b) austenite, insert – temperature dependence at 32 GHz.

*3.1 Field-cooling and measurements with H parallel to x axis*

First, the field-cooling and measurements were done with field along the longest *x*-axis. It was expected that after the transition from austenite to martensite, at about 45°C, in magnetic field, the single twin variant, with its' c-axis parallel to the field direction (*x*-variant) is formed. After the field-cooling, the resonance curves at temperature of 25 °C, were measured three times at a frequency of 32 GHz. The curves are identical with a strong single peak at $H_{res}$ = 2.41 kOe. This indicates that the sample remains in a single variant state when sweeping the applied field in the same direction as during the field-cooling. Then the resonance curves were measured at 25 °C for different frequencies from 4 to 42 GHz. The temperature was stepwise increased up to 120 °C and for each temperature a series of frequency measurements were done. The resonance fields as function of frequency for different temperatures are shown in Fig. 3. Because the martensitic alloy exhibits uniaxial magnetic anisotropy [16], the resonance field *H* should satisfy the modified Kittel's resonance condition for a thin film [17]:

$$(\omega/\gamma)^2 = (H_i + H_K) \cdot (H_i + H_K + 4\pi M_s), \quad (1)$$

where $H_i = H - 4\pi D_x M_s$ is the internal static field and $H_K$ is the effective anisotropy field. $D_x$ is the bulk demagnetizing factor of the sample in the direction of applied field, $\omega$ is the angular frequency and $\gamma$ is the gyromagnetic ratio. Using the spectroscopic factor *g* = 2.0 and saturation magnetization $M_s$, determined from VSM measurement, the anisotropy field $H_K$ can be calculated by fitting the measured resonance fields. The theoretically calculated resonance fields with corresponding values of $H_K$ (see Table 1) are shown by the full lines in Fig. 3a. In the martensitic state the FMR resonance peak is observed only for frequencies larger than 22 GHz. It is due to the strong uniaxial anisotropy with easy axis parallel to the field.

At the transition from martensitic to austenitic state (between 46 and 48°C) the resonance field sharply increases, as is illustrated for frequency 32 GHz in the insert of Fig. 3b. In austenite the single resonance peak is observed for all the frequencies. Even though the anisotropy of austenite is very weak [16] the frequency dependences of resonance fields cannot be fitted with $H_K = 0$. The effective anisotropy field decreases from about 0.4 kOe to zero when the temperature approaches Curie point.

Table 1. Fitting parameters for different temperatures and FMR measurement series: saturation magnetization, $4\pi M_s$, as defined from the hysteresis loops measured in VSM and effective anisotropy, $H_K$.

| Field during cooling | | $H \parallel x$ | | $H \parallel y$ | |
|---|---|---|---|---|---|
| Field during measurement | | $H \parallel x$ | $H \parallel y$ | $H \parallel y$ | $H \parallel x$ |
| T (°C) | $4\pi M_s$ (kGauss) | $H_K$ (kOe) | $H_K$ (kOe) | $H_K$ (kOe) | $H_K$ (kOe) |
| 25 | 7.129 | 7.169 | 7.168 | 7.148 | — |
| 30 | 7.018 | 6.940 | — | — | — |
| 35 | 6.901 | 6.679 | 6.694 | — | — |
| 40 | 6.771 | 6.427 | 6.444 | 6.442 | 6.352 |
| 45 | 6.621 | 6.111 | — | — | — |
| 50 | 5.778 | 0.389 | 0.389 | 0.413 | 0.389 |
| 60 | 5.350 | 0.346 | 0.359 | 0.376 | 0.346 |
| 80 | 4.240 | 0.231 | 0.226 | — | — |
| 100 | 2.497 | 0.074 | 0.026 | — | — |
| 120 | 0.751 | 0 | 0 | — | — |

*3.2 FMR measurements along y direction after field-cooling along x axis*

After the series of measurements with $H$ parallel to x-axis the sample was turned by 90° about z-axis and a similar series of measurements was performed with $H$ along y-axis. First, the frequency of 32 GHz was chosen and the measurements up to 27 kOe were repeated three times. In the first run some asymmetric signals were observed at low fields. However, in the second and third run a regular single peak at $H_{res}$ = 2.795 kOe appeared. This means that during the first sweep of magnetic field up to 27 kOe the magnetic reorientation of twin variants happened and the twin variants with their short easy c-axis parallel to the external field grew at the expense of the previously dominant x-variant. The resonance fields measured in this configuration are shown in Fig.4 and the fitted values of $H_K$ are shown in the fourth column of Table 1. The experimental results are similar to those described previously and shown in Fig.3. The resonances at all temperatures are now shifted to higher fields due to the higher demagnetizing factor ($D_y$) along y-axis of the sample.

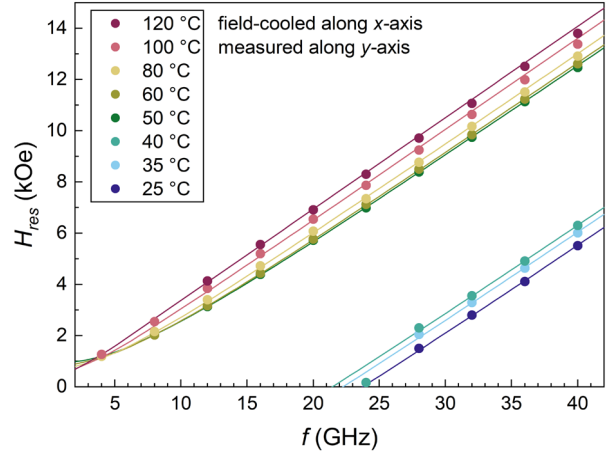

Fig. 4. Resonance fields as functions of frequency measured at different temperatures. Circles are the measured values; the lines represent theoretical fits. Before the measurements the sample was cooled in field parallel to *x*-axis. During the measurements magnetic field was applied along *y*-axis of the sample.

*3.3 Field-cooling and measurements with H parallel to y axis*

In the third type of measurements, the sample was cooled in field parallel to *y*-axis. Measurements with the field applied in the same direction and temperatures rising to 60 °C gave similar results as in the first series of measurements. The resonance fields shift to higher values is explained by the larger demagnetizing factor $D_y$. The fitting parameters $H_K$ are given in fifth column of Table 1.

*3.4 FMR measurements along x direction after field-cooling along y axis*

Finally, the sample was turned by 90 degrees and measurements with field parallel to *x*-axis were done. At the temperature of 25 °C the resonance peak of *x*-variant with c axis parallel to field does not exist. Instead of that a single peak at higher fields is observed for frequencies higher than 16 GHz (see Fig. 5a).

This peak can be attributed to the remnant of *y*-variant induced by the field cooling. Similar resonance spectra are obtained for 35 °C but with much weaker intensity (Fig. 5b). When the temperature increases to 40 °C these resonance peaks disappear and the standard resonances for single *x*-variant appear at lower fields (Fig. 5c). This is caused by the field induced reorientation of the martensitic twin structure taking place at 40 °C. Above the phase transition the resonance curves correspond to the austenite (Fig. 5d). The resonance fields for 5 different temperatures are shown on Fig. 6. The full lines denote the theoretically calculated resonance fields according to the modified Kittel's resonance conditions for single variants (Eqs. 1 and 2) with c-axis along the *x* or *y* sample axis, respectively. As can be seen the resonance fields of *y*-variant (black and cyan points in Fig. 6) do not correspond to the theoretical curves calculated from eq. 2 for a single variant with easy axis perpendicular to the field direction [17]:

$$(\omega/\gamma)^2 = (H_i - H_K) \cdot (H_i + 4\pi M_s), \qquad (2)$$

This indicates that at temperatures 25 °C and 35 °C the sample is not in a single variant state. The resonance fields are shifted to lower values due to a possible interaction with *x*-variant, which may be partially present. The resonance peaks of *x*-variant, however, are not visible because they are probably broad and may be overlapped by the noise. To explain the shift of resonance fields of the remnant *y*-variant at 25 °C and 35 °C a hypothetical effective field $h_y$ was introduced (see *Supplementary data*). The values of $h_y$ fitting the experimental points are shown in the lower part of Fig. 6.

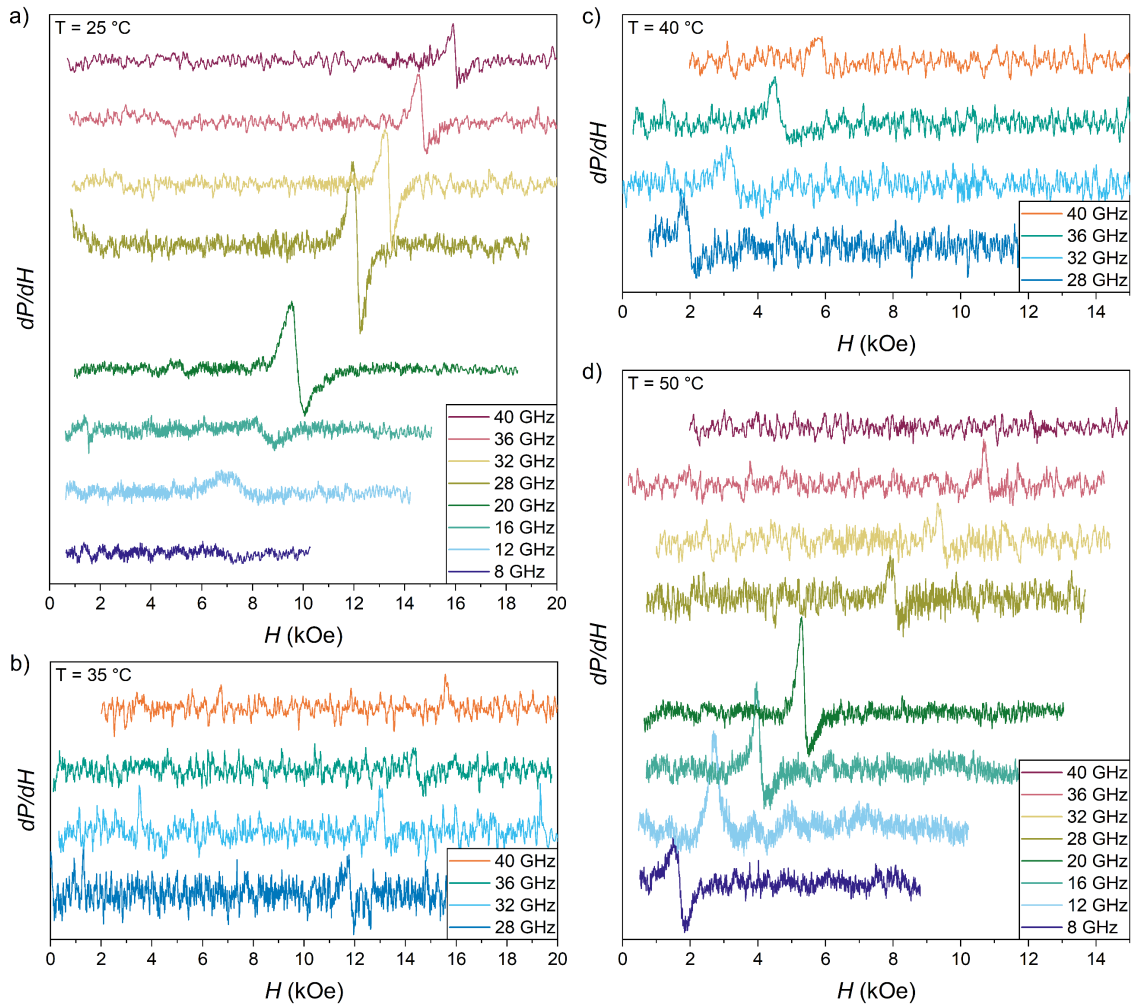

Fig. 5. Resonance curves measured with *H* parallel to *x*-axis at 4 different temperatures after previous field cooling and measurements with *H* parallel to *y*-axis: a) 25 °C, b) 35 °C, c) 40 °C, d) 50 °C.

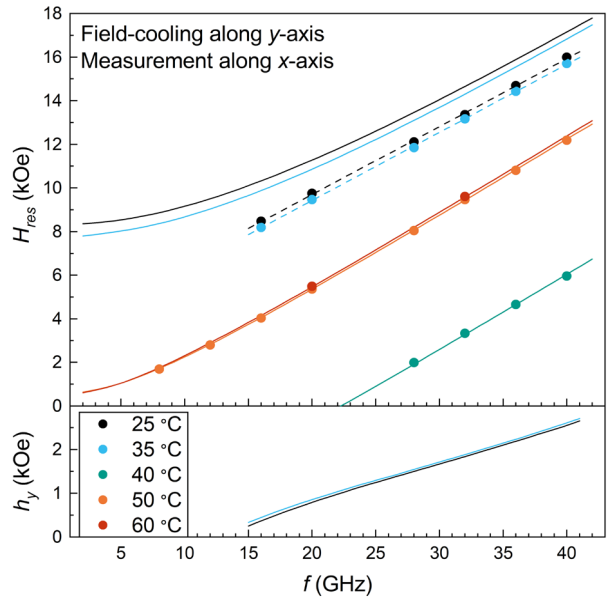

Fig. 6. Resonance fields as functions of frequency measured at different temperatures. Circles are the measured values; the lines represent theoretical fits. Before the measurements the sample was cooled in field parallel to *y*-axis. During the measurements, the magnetic field was applied along *x*-axis of the sample. The dotted black and cyan curves were obtained by introducing the effective field $h_y$ for the temperatures of 25 °C and 35 °C, presented in the lower part of the figure.

## 4. Discussion

When the sample is field-cooled in one of the <100> directions, of the austenite phase, a single variant martensite with its' c-axis along this direction is obtained. When the measuring field is applied in the same direction, the single variant state does not change. If, however, field is applied in a different direction, a magnetic reorientation takes place and a variant with its' c-axis closest to the field direction is formed. If the measuring field is along other <100> direction of the parental austenite, a new single variant martensitic state should be obtained. Whether the reorientation of crystal structure is complete depends on the magnitude of applied magnetic field and the mobility of twin boundaries. When the sample was field-cooled in $x$ direction and measured in $y$ direction, the complete reorientation was obtained after the first field sweep up to 27 kOe. While if the field-cooling was performed along $y$-axis and measurement along $x$-axis the field reorientation was not finished and both variants $x$ and $y$ coexisted during the measurements at temperatures 25 °C and 35 °C. This indicates that the mobility of twin boundaries was lower than in the previous case. Only when the temperature was increased to 40 °C the field reorientation took place and the resonance of the dominant $x$-variant at lower fields appeared. The reorientation of martensite at higher temperature can be explained by an increase of twin boundary mobility when the temperature approaches the transition temperature. It is, however, not clear why the mobility of twin boundaries is different for the two different orientations of magnetic field. It may be attributed to some small gradient of composition during the crystal growth.

The frequency dependence of $H_{res}$ for the single variant martensite well satisfies the modified Kittel's resonance condition (Eq. 1) with $g = 2$ and $H_K$ as shown in Table 1. The temperature dependence of anisotropy field $H_K$ is shown in Fig. 7. The estimated error of the measured resonance field is about 10% of the resonance linewidth (about 200 Oe). The same error can be expected for $H_K$. With the exception of two values the difference between $H_K$ obtained from the four types of measurements does not exceed much the estimated error. The anisotropy field measured by FMR at 25 °C is about 3% higher than 6.95 kOe determined from the room temperature VMS measurement of hysteresis loop along the hard anisotropy axis. At the transition to austenitic state $H_K$ sharply decreases and reaches zero at Curie temperature. The small anisotropy of austenite can be explained by a cubic anisotropy of premartensite [20,21]. The value $H_K = 0.39$ kOe just above the transition corresponds to the anisotropy constant $K_1$ about $9 \times 10^4$ erg/cm$^3$. This is comparable to the value $6 \times 10^4$ erg/cm$^3$ found by Tickle and James [20], but about one order larger than the anisotropy of nearly stochiometric $Ni_2MnGa$ at room temperature [22].

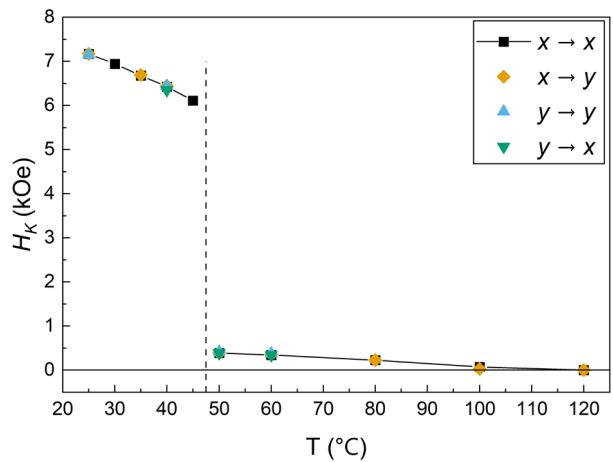

Fig. 7. Anisotropy field $H_K$ as a function of temperature.

Ferromagnetic resonance in the two-variant martensite obtained after field cooling along $y$ direction and field applied in $x$ direction (see Fig. 5) is more complicated. The resonance peaks at temperatures 25 and 35°C do not correspond to resonance fields for a single $y$-variant. The shift of $H_{res}$ to lower values can be attributed to the interaction between the two variants. To explain this effect, we introduce an effective field $h_y$ which hinders the magnetization of *y*-variant to rotate towards the direction of applied magnetic field. The origin of this field can be the dipolar and/or exchange coupling between the variants. Now the measured resonance fields can be fitted using appropriate values of $h_y$. The theoretical resonance fields are shown by the dashed lines in the upper part of Fig. 6 and the frequency dependence of appropriate field $h_y$ in the lower part of Fig. 6. The details of this fitting procedure are described in the *Supplementary material*. This is only a hypothetical explanation. The nature of the coupling cannot be distinguished from these measurements. It would require further investigation.

## 5. Conclusions

Broadband ferromagnetic resonance in single crystalline Ni-Mn-Ga shows that the large change of the magnetocrystalline anisotropy at the martensitic phase transition results in a sharp change of resonance field. In the single variant martensite phase the resonance field satisfies well the modified Kittel's resonance condition for a thin film with the splitting factor g = 2.0. If magnetic field is applied in parallel to the easy c-axis, the resonance is observed only for frequencies larger than 22 GHz. When the field is applied in a perpendicular direction, magnetic field induced reorientation of martensite structure takes place and resonance field changes. In the case of multivariant state the coupling between variants must be taken into account for the calculation of the resonance condition. Above the reverse martensite transformation temperature a weak magnetocrystalline anisotropy, comparable to previous reports, is observed. The exchange (optical) resonance modes reported previously [17] have not been observed.

## Acknowledgements

The authors acknowledge the assistance provided by the Ferroic Multifunctionalities project, supported by the Ministry of Education, Youth, and Sports of the Czech Republic. Project No. CZ.02.01.01/00/22_008/0004591, co-funded by the European Union. We acknowledge the support of the Czech Science Foundation (grant No. 22-22063S).

## Declaration of competing interest

The authors declare that they have no known competing financial interests or personal relationships that could have appeared to influence the work reported in this paper.

## Appendix A. Supplementary data

## Data availability

The data supporting the findings of this study is available at 10.5281/zenodo.15755284.

Appendix A. Supplementary data.

Fitting of resonance fields for temperatures 25 and 35 °C in Fig. 6.

Suppose that the sample contains two variants: *x*-variant with magnetization along the internal static field $H_i$ and *y*-variant perpendicular to it. If the twin boundary does not move the magnetization *M* of *y*-variant rotates towards the field direction.

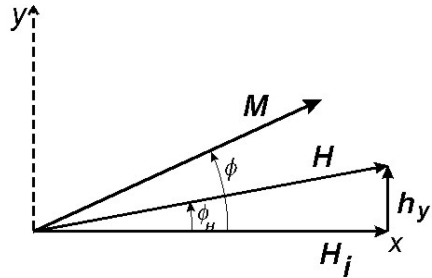

We suppose that the rotation of *y*-variant is impeded by some effective field $h_y$ due to interaction with *x*-variant. Then the total effective field acting on magnetization is $H = \sqrt{H_i^2 + h_y^2}$ which makes the angle $\phi_H$ with *x*-axis. The angle $\phi$ of magnetization *M* with respect to *x* direction can be obtained from the equilibrium condition

$$2H \sin(\phi - \phi_H) = H_K \sin 2\phi$$

Then the resonance condition for *y*-variant obtained from Smit-Beljers equation

$$(\omega/\gamma)^2 = \frac{E_{\theta\theta} E_{\phi\phi} - (E_{\theta\phi})^2}{M^2 \sin^2 \theta}$$

where $\theta$ is the angle of *M* with respect to *z*-axis, gives

$$(\omega/\gamma)^2 = [H \cos(\phi - \phi_H) + H_K \sin^2 \phi + 4\pi M_s][H \cos(\phi - \phi_H) - H_K \cos 2\phi]$$

The resonance field $H_{res}$ is then

$$H_{res} = H_i + 4\pi D_x M_s = H \cos \phi_H + 4\pi D_x M_s$$